\documentclass[12pt,a4paper]{article}

\usepackage{ifthen} 
\newboolean{pdflatex}
\setboolean{pdflatex}{true} 

\newboolean{articletitles}
\setboolean{articletitles}{true} 

\newboolean{uprightparticles}
\setboolean{uprightparticles}{false} 

\newboolean{inbibliography}
\setboolean{inbibliography}{false} 

\usepackage[top=1in, bottom=1.25in, left=1in, right=1in]{geometry}

\usepackage{microtype}
\usepackage{lineno}  
\usepackage{xspace} 
\usepackage{caption} 

\usepackage{graphicx}  
\usepackage{color}
\usepackage{colortbl}
\graphicspath{{./figs/}} 

\usepackage{amsmath} 
\usepackage{amssymb}
\usepackage{amsfonts}
\usepackage{upgreek} 

\newcommand*\patchAmsMathEnvironmentForLineno[1]{%
\expandafter\let\csname old#1\expandafter\endcsname\csname #1\endcsname
\expandafter\let\csname oldend#1\expandafter\endcsname\csname
end#1\endcsname
 \renewenvironment{#1}%
   {\linenomath\csname old#1\endcsname}%
   {\csname oldend#1\endcsname\endlinenomath}%
}
\newcommand*\patchBothAmsMathEnvironmentsForLineno[1]{%
  \patchAmsMathEnvironmentForLineno{#1}%
  \patchAmsMathEnvironmentForLineno{#1*}%
}
\AtBeginDocument{%
\patchBothAmsMathEnvironmentsForLineno{equation}%
\patchBothAmsMathEnvironmentsForLineno{align}%
\patchBothAmsMathEnvironmentsForLineno{flalign}%
\patchBothAmsMathEnvironmentsForLineno{alignat}%
\patchBothAmsMathEnvironmentsForLineno{gather}%
\patchBothAmsMathEnvironmentsForLineno{multline}%
\patchBothAmsMathEnvironmentsForLineno{eqnarray}%
}

\usepackage{hyperref}    
\usepackage[all]{hypcap} 

\usepackage{xspace} 
\usepackage{upgreek}

\def\lhcb {\mbox{LHCb}\xspace}

\def\MagUp {\mbox{\em Mag\kern -0.05em Up}\xspace}

\ifthenelse{\boolean{uprightparticles}}%
{

 \def\PDelta      {\ensuremath{\Delta}\xspace}                 
 \def\PXi      {\ensuremath{\Xi}\xspace}                 
 \def\PLambda      {\ensuremath{\Lambda}\xspace}                 
 \def\PSigma      {\ensuremath{\Sigma}\xspace}                 
 \def\POmega      {\ensuremath{\Omega}\xspace}                 
 \def\PUpsilon      {\ensuremath{\Upsilon}\xspace}                 
 
 \def\PB      {\ensuremath{\mathrm{B}}\xspace}                 
                  
 \def\PD      {\ensuremath{\mathrm{D}}\xspace}

 \def\PK      {\ensuremath{\mathrm{K}}\xspace}

 \def\Pi      {\ensuremath{\mathrm{i}}\xspace}

 \def\Pp      {\ensuremath{\mathrm{p}}\xspace}

}
{

 \mathchardef\PDelta="7101
 \mathchardef\PXi="7104
 \mathchardef\PLambda="7103
 \mathchardef\PSigma="7106
 \mathchardef\POmega="710A
 \mathchardef\PUpsilon="7107
                  
 \def\PB      {\ensuremath{B}\xspace}                 
                  
 \def\PD      {\ensuremath{D}\xspace}

 \def\PK      {\ensuremath{K}\xspace}

 \def\Pi      {\ensuremath{i}\xspace}

 \def\Pp      {\ensuremath{p}\xspace}

}

\makeatletter
\ifcase \@ptsize \relax
  \newcommand{\miniscule}{\@setfontsize\miniscule{4}{5}}
\or
  \newcommand{\miniscule}{\@setfontsize\miniscule{5}{6}}
\or
  \newcommand{\miniscule}{\@setfontsize\miniscule{5}{6}}
\fi
\makeatother

\DeclareRobustCommand{\optbar}[1]{\shortstack{{\miniscule (\rule[.5ex]{1.25em}{.18mm})}
  \\ [-.7ex] $#1$}}

  \def\Kbar    {{\kern 0.2em\overline{\kern -0.2em \PK}{}}\xspace}

\def\KorKbar    {\kern 0.18em\optbar{\kern -0.18em K}{}\xspace}

  \def\Dbar    {{\kern 0.2em\overline{\kern -0.2em \PD}{}}\xspace}

\def\DorDbar    {\kern 0.18em\optbar{\kern -0.18em D}{}\xspace}

\def\B       {{\ensuremath{\PB}}\xspace}
\def\Bbar    {{\ensuremath{\kern 0.18em\overline{\kern -0.18em \PB}{}}}\xspace}

\def\BorBbar    {\kern 0.18em\optbar{\kern -0.18em B}{}\xspace}

  \def\Y#1S{\ensuremath{\PUpsilon{(#1S)}}\xspace}

\def\proton      {{\ensuremath{\Pp}}\xspace}
\def\antiproton  {{\ensuremath{\overline \proton}}\xspace}

\def\Lbar        {{\ensuremath{\kern 0.1em\overline{\kern -0.1em\PLambda}}}\xspace}
\def\LorLbar    {\kern 0.18em\optbar{\kern -0.18em \PLambda}{}\xspace}

\def\to                 {\ensuremath{\rightarrow}\xspace}

\def\AT#1     {\ensuremath{A_{\mathrm{T}}^{#1}}\xspace}           

\def\C#1      {\ensuremath{\mathcal{C}_{#1}}\xspace}                       
\def\Cp#1     {\ensuremath{\mathcal{C}_{#1}^{'}}\xspace}                    
\def\Ceff#1   {\ensuremath{\mathcal{C}_{#1}^{\mathrm{(eff)}}}\xspace}        
\def\Cpeff#1  {\ensuremath{\mathcal{C}_{#1}^{'\mathrm{(eff)}}}\xspace}       
\def\Ope#1    {\ensuremath{\mathcal{O}_{#1}}\xspace}                       
\def\Opep#1   {\ensuremath{\mathcal{O}_{#1}^{'}}\xspace}                    

\newcommand{\tev}{\ifthenelse{\boolean{inbibliography}}{\ensuremath{~T\kern -0.05em eV}\xspace}{\ensuremath{\mathrm{\,Te\kern -0.1em V}}}\xspace}
\newcommand{\gev}{\ensuremath{\mathrm{\,Ge\kern -0.1em V}}\xspace}
\newcommand{\mev}{\ensuremath{\mathrm{\,Me\kern -0.1em V}}\xspace}
\newcommand{\kev}{\ensuremath{\mathrm{\,ke\kern -0.1em V}}\xspace}
\newcommand{\ev}{\ensuremath{\mathrm{\,e\kern -0.1em V}}\xspace}
\newcommand{\gevc}{\ensuremath{{\mathrm{\,Ge\kern -0.1em V\!/}c}}\xspace}
\newcommand{\mevc}{\ensuremath{{\mathrm{\,Me\kern -0.1em V\!/}c}}\xspace}
\newcommand{\gevcc}{\ensuremath{{\mathrm{\,Ge\kern -0.1em V\!/}c^2}}\xspace}
\newcommand{\gevgevcccc}{\ensuremath{{\mathrm{\,Ge\kern -0.1em V^2\!/}c^4}}\xspace}
\newcommand{\mevcc}{\ensuremath{{\mathrm{\,Me\kern -0.1em V\!/}c^2}}\xspace}

\def\invfb   {\ensuremath{\mbox{\,fb}^{-1}}\xspace}

\def\sec  {\ensuremath{\mathrm{{\,s}}}\xspace}

\def\ps   {\ensuremath{{\mathrm{ \,ps}}}\xspace}

\def\gsim{{~\raise.15em\hbox{$>$}\kern-.85em
          \lower.35em\hbox{$\sim$}~}\xspace}
\def\lsim{{~\raise.15em\hbox{$<$}\kern-.85em
          \lower.35em\hbox{$\sim$}~}\xspace}

\def\sPlot{\mbox{\em sPlot}\xspace}

\def\pt         {\mbox{$p_{\mathrm{ T}}$}\xspace}

\def\evtgen     {\mbox{\textsc{EvtGen}}\xspace}

\def\geant      {\mbox{\textsc{Geant4}}\xspace}

\def\photos     {\mbox{\textsc{Photos}}\xspace}

\def\pythia     {\mbox{\textsc{Pythia}}\xspace}

\def\tell1  {TELL1\xspace}
\def\ukl1   {UKL1\xspace}

\def\Sigmaplus{\ensuremath{\PSigma^+}\xspace}
\def\Sigmaplusbar{\ensuremath{\overline{\PSigma^+}}\xspace}

\def\sigmapmumu{\ensuremath{\PSigma^+ \to \Pp \mu^+ \mu^-}\xspace}
\def\pmumu{\ensuremath{\Pp \mu^+ \mu^-}\xspace}
\def\mpmumu{\ensuremath{m_{\Pp \mu^+ \mu^-}}\xspace}

\def\sigmapxmumu{\ensuremath{\PSigma^+ \to \Pp X^0 (\to \mu^+ \mu^-)}\xspace}

\def\pmumulfv{\ensuremath{\antiproton \mu^+ \mu^+}\xspace}
\def\sigmappiz{\ensuremath{\PSigma^+ \to \Pp \pi^0}\xspace}
\def\sigmappizero{\sigmappiz}

\def\pizero{\ensuremath{\pi^0}\xspace}

\def\kpipipi{\ensuremath{K^+ \to \pi^+ \pi^- \pi^+}\xspace}
\def\kpimumu{\ensuremath{K^+ \to \pi^+ \mu^- \mu^+}\xspace}

\def\mmumu{\ensuremath{m_{\mu^+\mu^-}}\xspace}
\def\lambdappi{\ensuremath{\PLambda \to \Pp \pi^-}\xspace}
\def\B{\ensuremath{\mathcal{B}}\xspace}

\def\bujpsikstar{\ensuremath{B^+ \to J/\psi K^{\ast +}(\to K^+ \pi^0) }\xspace}
\def\bujpsik{\ensuremath{B^+ \to J/\psi K^+}\xspace}

\usepackage{booktabs}

\usepackage{xcolor} 
\definecolor{halfgray}{gray}{0.55} 
\definecolor{webgreen}{rgb}{0,.5,0}
\definecolor{webbrown}{rgb}{.6,0,0}
\definecolor{Maroon}{cmyk}{0, 0.87, 0.68, 0.32}
\definecolor{RoyalBlue}{cmyk}{1, 0.50, 0, 0}

\hypersetup{%
    colorlinks=true, linktocpage=true, pdfstartpage=3, pdfstartview=FitV,%
    breaklinks=true, pdfpagemode=UseNone, pageanchor=true, pdfpagemode=UseOutlines,%
    plainpages=false, bookmarksnumbered, bookmarksopen=true, bookmarksopenlevel=1,%
    urlcolor=webbrown, linkcolor=RoyalBlue, citecolor=webgreen, 
    pdftitle={Evidence for the rare decay Sigma+ -> p mu+ mu-  },%
    pdfauthor={Francesco Dettori},%
    pdfsubject={},%
    pdfkeywords={},%
    pdfcreator={pdfLaTeX},%
  }

\usepackage{cite} 
\usepackage{mciteplus}
\usepackage{overpic}
\usepackage{tikz}
\usetikzlibrary{decorations.pathreplacing,decorations.pathmorphing,
decorations.markings,trees,calc,patterns}
\usetikzlibrary{positioning,arrows}
\usetikzlibrary{decorations.pathmorphing}
\usetikzlibrary{decorations.markings}

\tikzset{
photon/.style={decorate, decoration={snake}, draw=webbrown},
higgs/.style={decorate, dashed, draw=webbrown},
particle/.style={draw=RoyalBlue, postaction={decorate},decoration={markings,mark=at 
position .5 with {\arrow[draw=RoyalBlue]{>}}}},
antiparticle/.style={draw=RoyalBlue, 
postaction={decorate},decoration={markings,mark=at position .5 with 
{\arrow[draw=RoyalBlue]{<}}}}, 
gluon/.style={decorate, draw=black,decoration={snake,amplitude=4pt, segment 
length=5pt}}, 
majorana/.style={draw=black, postaction={decorate},decoration={markings,mark=at 
position .48 with {\arrow[draw=black]{>}},mark=at position .52 with 
{\arrow[draw=black]{<}}}},
gluonloop/.style={circle, decorate, draw=black, 
decoration={coil,aspect=1.2,amplitude=2pt, segment length=4pt},minimum 
height=1.2em},
}

\def\NSigmappizero{\ensuremath{(1171 \pm 9)\times 10^{3}}\xspace}
\def\NTOTSigma{\ensuremath{ 10^{14}}\xspace}

\def\alphafull{\ensuremath{(2.2 \pm 1.2)\times 10^{-9}}\xspace}
\def\expevents{\ensuremath{23 \pm 20}\xspace}
\def\signdefault{\ensuremath{4.1}\xspace}
\def\nsigmadefault{\ensuremath{ 10.2\,^{+\,3.9}_{-\,3.5}}\xspace}

\def\upperlimithcpninety{\ensuremath{1.4 \times 10^{-8}}\xspace}
\def\upperlimithcpninetyfive{\ensuremath{1.7 \times 10^{-8}}\xspace}

\def\candidateshcp{\ensuremath{3.0\,^{+\,1.7}_{-\,1.4}}\xspace}
\def\fractionhcp{\ensuremath{30\%}\xspace}

\def\brmeasured{\ensuremath{(2.2\,^{+\,1.8}_{-\,1.3})\times 10^{-8}}\xspace}
\def\brmeasuredsyst{\ensuremath{(2.2\,^{+\,0.9}_{-\,0.8}\,^{+\,1.5}_{-\,1.1})\times 10^{-8}}\xspace}

\usepackage{longtable} 

\def\myfigurewidth{0.6\textwidth}
\begin{document}

\renewcommand{\thefootnote}{\fnsymbol{footnote}}
\setcounter{footnote}{1}


\begin{titlepage}
\pagenumbering{roman}

\vspace*{-1.5cm}
\centerline{\large EUROPEAN ORGANIZATION FOR NUCLEAR RESEARCH (CERN)}
\vspace*{1.5cm}
\noindent
\begin{tabular*}{\linewidth}{lc@{\extracolsep{\fill}}r@{\extracolsep{0pt}}}
\ifthenelse{\boolean{pdflatex}}
{\vspace*{-2.7cm}\mbox{\!\!\!\includegraphics[width=.14\textwidth]{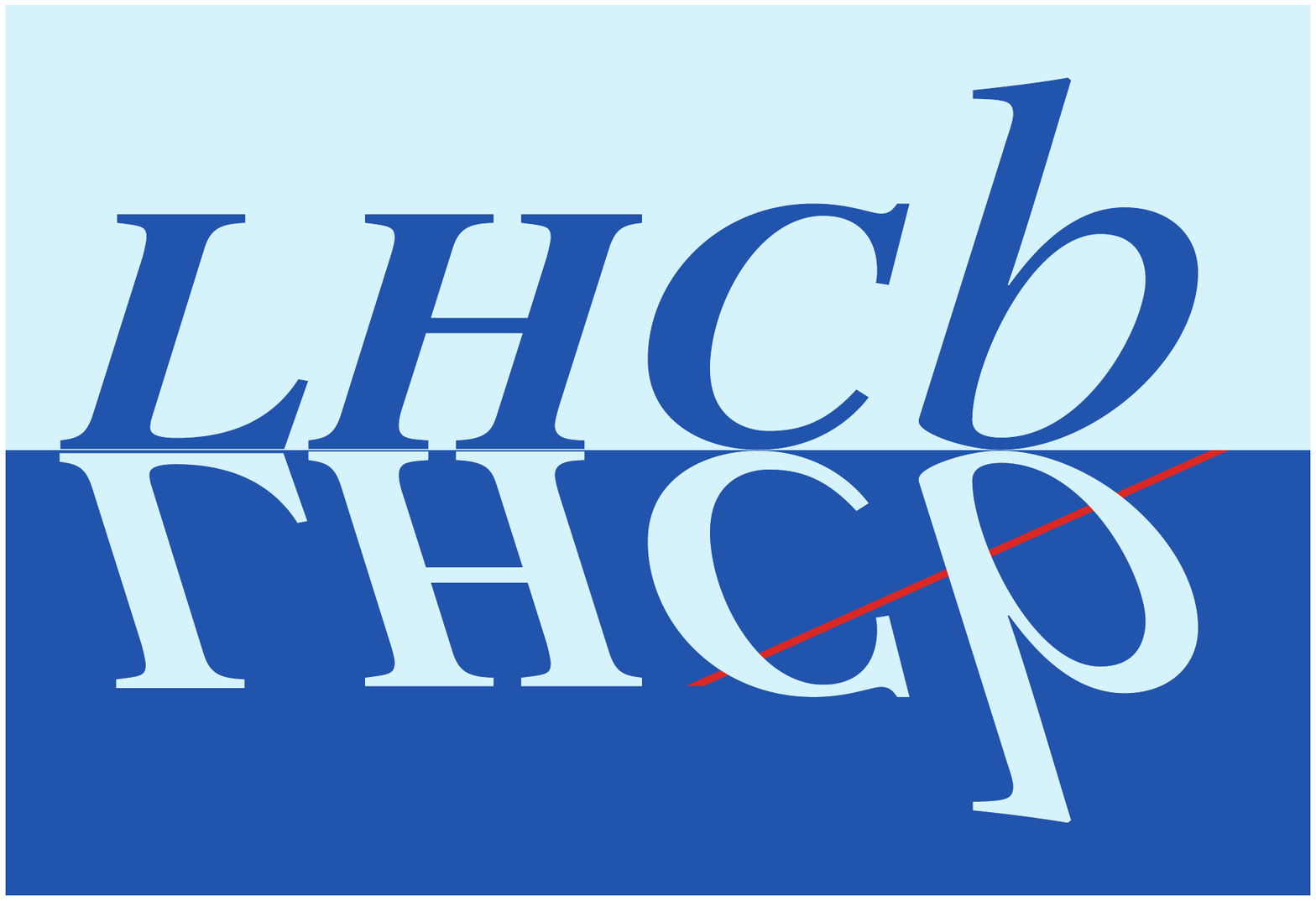}} & &}%
{\vspace*{-1.2cm}\mbox{\!\!\!\includegraphics[width=.12\textwidth]{lhcb-logo.eps}} & &}%
\\
 & & CERN-EP-2017-319 \\  
 & & LHCb-PAPER-2017-049 \\  
 & & \today \\ 
 & & \\
\end{tabular*}

\vspace*{4.0cm}

{\normalfont\bfseries\boldmath\huge
\begin{center}
    Evidence for the rare decay \sigmapmumu  
\end{center}
}

\vspace*{2.0cm}

\begin{center}
The LHCb collaboration\footnote{Authors are listed at the end of this Letter.}
\end{center}

\vspace{\fill}

\begin{abstract}
  \noindent
A search for the rare decay \sigmapmumu is performed using $pp$ collision data recorded by the LHCb experiment
at centre-of-mass energies $\sqrt{s} = 7$ and $8 \tev$, corresponding to an integrated luminosity of $3 \invfb$. 
An excess of events is observed with respect to the background expectation, with a signal significance 
of \signdefault standard deviations.  No significant structure is observed in the dimuon invariant mass distribution, 
in contrast with a previous result from the HyperCP experiment.
The measured \sigmapmumu branching fraction is \brmeasured, where statistical and systematic uncertainties are included,
which is consistent with the Standard Model prediction.
\end{abstract}

\vspace*{2.0cm}

\begin{center}
  Published in Phys. Rev. Lett. 120 (2018) 221803.
\end{center}

\vspace{\fill}

{\footnotesize 
\centerline{\copyright~CERN on behalf of the \lhcb collaboration, licence \href{http://creativecommons.org/licenses/by/4.0/}{CC-BY-4.0}.}}
\vspace*{2mm}

\end{titlepage}


\newpage
\setcounter{page}{2}
\mbox{~}
%
%
%
%

\cleardoublepage


\renewcommand{\thefootnote}{\arabic{footnote}}
\setcounter{footnote}{0}



\pagestyle{plain} 
\setcounter{page}{1}
\pagenumbering{arabic}


The \sigmapmumu decay
is an $s\to d$ quark-flavour-changing neutral-current process, allowed only at loop level in the
standard model (SM).
The process is dominated by long-distance contributions for a predicted
branching fraction of 
$ 1.6 \times 10^{-8} < \mathcal{B}(\sigmapmumu) < 9.0 \times
10^{-8}$~\cite{He:2005yn}, 
while the short-distance SM contributions are suppressed and contribute to the branching fraction
at the level of about $10^{-12}$.
Evidence for this decay was reported by the HyperCP collaboration~\cite{Park:2005eka}
with a measured branching fraction 
$\mathcal{B}(\sigmapmumu) = (8.6\,^{+\,6.6}_{-\,5.4} \pm 5.5) \times 10^{-8}$,  
which is compatible with the SM prediction.
HyperCP observed three candidates; remarkably, all of them have almost the same dimuon invariant
mass of $m_{X^0}  = 214.3 \pm 0.5 \mevcc$, close to the lower kinematic limit.
Such a distribution, if confirmed, would point towards a process 
with an intermediate particle $X^0$ coming from the \Sigmaplus baryon and decaying
into 
two muons, \emph{i.e.} a \sigmapxmumu decay, which would constitute evidence for
physics beyond the SM~(BSM). 
Various BSM theories have been proposed to explain the HyperCP result. The
intermediate $X^0$ particle 
could be, for example, a light pseudoscalar Higgs boson~\cite{He:2006fr,He:2006uu}
or a sgoldstino~\cite{Gorbunov:2005nu,Demidov:2006pt} in various supersymmetric
models. Other interpretations and implications 
can be found in
Refs.~\cite{He:2005we,Geng:2005ra,Deshpande:2005mb,Chen:2007uv,Xiangdong:2007vv,
Mangano:2007gi,Pospelov:2008zw}; 
in general a pseudoscalar particle is favoured over a scalar particle and a lifetime
of the order of $10^{-14}\sec$ is estimated for the former case.
Attempts to confirm the existence of this $X^0$ particle have been made at
several experiments in various initial and final states
without finding any signal
\cite{Love:2008aa,Tung:2008gd,Abazov:2009yi,Aubert:2009cp,Hyun:2010an,
Abouzaid:2011mi,Lees:2014xha,Ablikim:2015voa}; 
these null results include studies of the decays $B^{0}_{(s)}\to
\mu^+\mu^-\mu^+\mu^-$~\cite{Aaij:2013lla},  $B^0\to K^{\ast
0}\mu^+\mu^-$~\cite{Aaij:2015tna}, 
$B^+\to K^+ \mu^+\mu^-$~\cite{LHCb-PAPER-2016-052}, 
and a search for photon-like particles~\cite{LHCb-PAPER-2017-038} by the LHCb experiment. 
However, the search for the \sigmapmumu decay has not been repeated
 due to the lack of experiments with large hyperon production rates and 
to the experimental difficulty of reconstructing soft and long-lived hadrons. 

Hyperons are produced copiously in high-energy proton-proton collisions at the
Large Hadron Collider.
A search for \sigmapmumu decays at the LHCb experiment, as also suggested in 
Ref.~\cite{Park:2010zze}, could therefore confirm
or disprove the HyperCP evidence, and the branching fraction can be measured.
This Letter presents a search for the \sigmapmumu decay performed using $pp$ collision data 
recorded by the LHCb experiment at centre-of-mass energies $\sqrt{s} = 7$ and $8
\tev$, corresponding to an integrated luminosity of $3 \invfb$.
The inclusion of charge-conjugated processes is implied throughout this Letter.

This search follows a strategy similar to that of other studies of rare decays in LHCb, 
although with differences due to the relatively low transverse momenta of the final-state particles.
First, a loose selection is applied based on geometric and kinematic variables. 
The final sample is obtained rejecting the background with requirements 
on the output of a multivariate selection, based 
on a boosted decision tree algorithm~(BDT)~\cite{Breiman,AdaBoost},
and on particle identification variables. 
The signal yield is obtained from a fit to the \pmumu invariant-mass spectrum and is converted 
into a branching fraction by normalising to the \sigmappizero control channel. 
The analysis is designed in order to search for possible peaks in the dimuon
invariant-mass distribution, in view of the possible existence of unknown intermediate particles. 

The \lhcb detector is a single-arm forward
spectrometer covering the \mbox{pseudorapidity} range $2<\eta <5$,
described in detail in Refs.~\cite{Alves:2008zz,LHCb-DP-2014-002}. 
It includes a high-precision tracking system
consisting of a silicon-strip vertex detector surrounding the $pp$
interaction region, a large-area silicon-strip detector located
upstream of a dipole magnet with a bending power of about
$4{\mathrm{\,Tm}}$, and three stations of silicon-strip detectors and straw
drift tubes placed downstream of the magnet.
Particle identification is provided by two ring-imaging Cherenkov detectors, 
an electromagnetic and a hadronic calorimeter, and a muon system 
composed of alternating layers of iron and multiwire proportional chambers.

The online event selection is performed by a trigger system, 
which consists of a hardware stage, based on information from the calorimeter and muon
systems, followed by two software stages. The first software stage performs a preliminary event reconstruction
based on partial information while the second applies a full event reconstruction.
Each of the three trigger stages is divided into many trigger selections dedicated to various types of signal.
The final-state particles from the signal decay involved in this analysis 
typically have insufficient transverse momenta to satisfy the requirements of one or more trigger stages.
Nevertheless, given the large production rate of \Sigmaplus baryons in $pp$ collisions, 
the present search can be performed with data selected at one or more trigger stages by other particles in the event.
In the offline processing, trigger decisions are associated with reconstructed candidates.
A trigger decision can thus be ascribed to the reconstructed candidate, the rest of the event or a combination of both;
events triggered as such are defined respectively as triggered on signal (TOS), triggered independently of signal (TIS), and triggered on both.
While all the candidates passing the trigger selection are used in the search for \sigmapmumu decays, 
only the TIS candidates are used in the normalisation channel \sigmappizero.
Furthermore, control channels with large yields are exploited to estimate the trigger efficiency 
by measuring the overlap of candidates which are TIS and TOS simultaneously~\cite{LHCb-DP-2012-004}. 

Simulation is used to devise and optimise the analysis strategy, 
as well as to estimate reconstruction and selection efficiencies. 
In the simulation, $pp$ collisions are generated using \pythia~\cite{Sjostrand:2006za,*Sjostrand:2007gs} 
with a specific \lhcb configuration~\cite{LHCb-PROC-2010-056}.  Decays of hadronic particles
are described by \evtgen~\cite{Lange:2001uf}, in which final-state
radiation is generated using \photos~\cite{Golonka:2005pn}. The
interaction of the generated particles with the detector, and its response,
are implemented using the \geant toolkit~\cite{Allison:2006ve, *Agostinelli:2002hh}, as described in
Ref.~\cite{LHCb-PROC-2011-006}.
The signal \sigmapmumu decay is generated according to a phase-space model.


Candidate \sigmapmumu decays are selected by combining two good-quality 
oppositely charged tracks identified as muons with a third track identified as a proton.
The three tracks are required to form a secondary vertex (SV) with a good vertex-fit quality. 
The short lifetime estimated for the $X^0$ particle would result in a prompt signal in this search, 
hence no attempt is made to distinguish the dimuon origin vertex from the SV of the \Sigmaplus baryon.
The measured \Sigmaplus candidate proper decay time 
is required to be greater than $6 \ps$, ensuring that the SV is displaced from any 
$pp$ interaction vertex (primary vertex, PV). 
The final-state particles are required to be inconsistent with originating from any PV in the event.
Only \Sigmaplus candidates with transverse momentum $\pt > 0.5 \gevc$  
and a decay topology consistent with a particle originating from the PV are retained. 
A candidate \sigmapmumu decay is considered only if its invariant mass, \mpmumu, satisfies 
$|\mpmumu - m_{\Sigmaplus}| < 500\mevcc$, 
where $m_{\Sigmaplus}$ is the known mass of the \Sigmaplus particle~\cite{PDG2016}.
The background component due to \lambdappi decays is 
vetoed by discarding candidates having a $p\mu^-$ pair invariant mass, 
calculated with the $p \pi^-$ mass hypothesis,  within 10\mevcc from the known $\PLambda$ mass~\cite{PDG2016}.
Possible backgrounds from decays peaking in the \pmumu invariant mass 
have been examined, including \kpipipi, \kpimumu, and various hyperon 
decays, and none has been found to contribute significantly. 
After all selection requirements, no retained event contains more than one candidate.

Candidate \sigmappizero decays are selected by combining one good-quality track 
identified as a proton with a \pizero reconstructed in the $\pizero \to \gamma \gamma$ mode 
from two clusters in the electromagnetic calorimeter. 
Given the impossibility to reconstruct the $\Sigma^+$ decay SV with the proton track only, 
the momentum direction of the $\pi^0$ is calculated assuming the $\pi^0$ is produced at the PV.
The selection of \sigmappizero decays is similar to that of the signal, with 
tighter requirements applied, in order to reduce the large combinatorial background,
on the proton identification and on the transverse momenta of the final-state particles
($\pt > 0.5 \gevc$  for the proton and $\pt>0.7 \gevc$ for the \pizero).
Finally, candidate \kpipipi decays, selected as control channel for various parts of 
the analysis, are required to pass a selection similar to that of the signal, 
starting from three good-quality tracks, with total charge equal to $\pm 1$, 
and which are assigned the pion mass hypothesis without requirements on 
the identification of the particle.

The sample of \sigmapmumu candidates in data after the initial selection is dominated by 
combinatorial background, part of which is due to misidentified particles.  
This background is rejected by placing requirements on the BDT output variable and on multivariate particle identification 
variables~\cite{LHCb-DP-2014-002} on the muons and on the proton. 
The BDT combines information from the following input variables: 
the angle between the \Sigmaplus reconstructed momentum and the vector joining the PV to the SV, 
the flight distance significance of the \Sigmaplus candidate, 
the distance of closest approach among the final-state particles, 
the transverse momenta of the final-state particles, 
the impact parameter $\chi^2$ ($\chi^2_{\rm IP}$) of the final-state particles, 
defined as the difference between the vertex-fit $\chi^2$ of a PV formed with and without 
the particle in question, the $\chi^2_{\rm IP}$ of the \Sigmaplus candidate, the $\chi^2$ of the SV, 
and an isolation variable constructed from the number of tracks within an 
angular cone around each of the final-state particles. 
These variables are chosen so that the dependence on the \pmumu 
invariant mass and on the dimuon invariant mass is small and linear to minimise 
potential biases.
The BDT is optimised using simulated samples of \sigmapmumu events for the 
signal and \pmumulfv candidates in data for the background.
The selection for the control \pmumulfv sample is identical to that of the 
signal but considering muons of identical charge. 
The final selection criteria are chosen in order to optimise the potential to obtain 
evidence for a signal with a branching fraction as small as possible~\cite{Punzi:2003bu}. 
No BDT selection is applied to the normalisation and control channels.

The number of signal candidates is converted into a branching fraction with the 
formula
\begin{eqnarray*}\label{eq:normsigma}
 \B(\sigmapmumu) &=&
\frac{\varepsilon_{\sigmappizero}}{\varepsilon_{\sigmapmumu}}
\frac{N_{\sigmapmumu}}{N_{\sigmappizero}} \B(\sigmappizero)\\
 &=& \alpha \cdot N_{\sigmapmumu}\quad,
\end{eqnarray*}
where $\varepsilon$, $N$ and $\B$ are the efficiency, candidate yield and 
branching fraction of the corresponding channel, respectively, and $\alpha$ is 
the single-event sensitivity.
The ratio of signal and normalisation channel efficiencies, which includes the 
acceptance, the trigger efficiency, the reconstruction efficiency of the final-state particles and the 
selection efficiency, is computed with samples of simulated events corrected to 
take into account known differences between data and simulation.  
The reconstruction efficiency for the \pizero is calibrated using the ratio of 
\bujpsikstar and \bujpsik decays reconstructed in data~\cite{Govorkova:2015vqa}.
The particle-identification efficiencies of protons and muons are calibrated 
with control channels in data. 
Residual differences between data and simulation are treated as sources of 
systematic uncertainty. 
The ratio of the trigger efficiencies for the signal and normalisation channels
is estimated with simulated samples and cross-checked in data: 
the trigger efficiency is obtained for selected trigger lines from the overlap of TIS and TOS events in the normalisation channel
and is compared between data and simulation~\cite{LHCb-DP-2012-004}.
The small size of this overlap induces a 40\% relative systematic uncertainty 
associated with the trigger efficiency ratio.  
The ratio of the trigger efficiencies is of the order of 0.09, 
owing to the use of all events for the signal, while TIS-only events are used for the normalisation channel. 
Possible differences in the BDT selection efficiency for the 
\sigmapmumu signal in data and in simulation are calibrated using the \kpipipi control channel. 
The sources of systematic uncertainties associated with the normalisation are reported in Table~\ref{tab:systematics}.

\begin{table}[tb]
\begin{center}
\caption{Relative systematic uncertainties associated with the normalisation. }\label{tab:systematics}
\begin{tabular}{lrrl}
\toprule 
   Source 	& Uncertainty \\
\midrule
   Selection efficiency		& 1\%  \\
   BDT efficiency 		& 6\% 	\\
   PID efficiency ratio		& 28\% 	\\
   \pizero efficiency 	& 10\%	 \\
   Trigger efficiency ratio		& 40\%   \\
   \midrule 
   Total 					&  50\% \\
   \bottomrule
  \end{tabular}               \end{center}
  \end{table}

\begin{figure}[tb]
\begin{center}
 \includegraphics[width = \myfigurewidth]{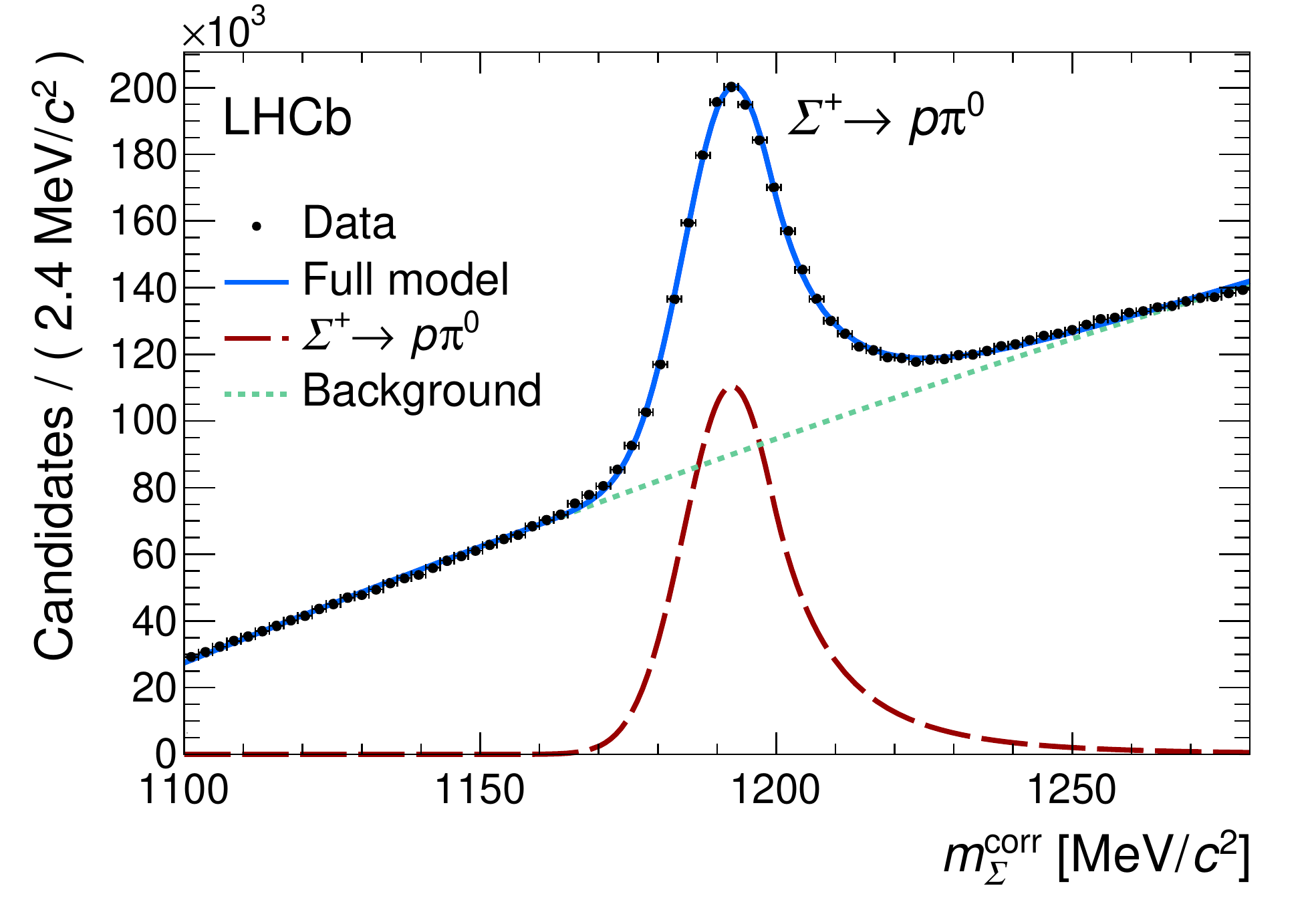}
 \caption{Distribution of the corrected mass 
$m_{\PSigma}^{\text{corr}}$, defined as in the text,  for \sigmappizero candidates superimposed with the 
fit to data.}
\label{fig:sigmappizero}
\end{center}
\end{figure}

The observed number of \sigmappizero candidates is \NSigmappizero, as obtained
from a binned extended maximum likelihood fit to the corrected invariant mass 
distribution $m_{\PSigma}^{\text{corr}}$.
The corrected invariant mass is defined as 
$m_{\PSigma}^{\text{corr}} = m_{p\gamma\gamma} - m_{\gamma\gamma} + m_{\pi^0}$,
where $m_{\pi^0}$ is the known mass of the $\pi^0$ meson~\cite{PDG2016},
to account for the limited precision in the reconstructed invariant mass of the two photons ($m_{\gamma\gamma}$). 
The \sigmappizero distribution is described as a Gaussian function with a power-law
tail on the higher-mass side, while the background is described by a modified ARGUS 
function~\cite{Albrecht:1990am}, where the power parameter is allowed 
to vary as in Ref.~\cite{Dobbs:2007ab}.
The distribution is shown in Fig.~\ref{fig:sigmappizero}, 
superimposed with the fit. 

The single-event sensitivity is $\alpha = \alphafull$, where the uncertainty is 
dominated by the systematic contribution. This sensitivity corresponds to 
about \NTOTSigma \Sigmaplus baryons produced in the LHCb acceptance in the 
considered dataset. The number of expected signal \sigmapmumu candidates is 
\expevents assuming a branching 
fraction of $(5\pm 4)\times 10^{-8}$, to cover the range predicted by the SM.

The observed number of signal \sigmapmumu decays is obtained with a fit to the 
\pmumu invariant-mass distribution in the range 
$1149.6 < \mpmumu < 1409.6 \mevcc$.
The signal distribution is described by an Hypatia function~\cite{Santos:2013gra}. 
The peak position and resolution are calibrated using the control channel 
\kpipipi and by comparing distributions in data and simulation. 
No bias is seen in the peak position, while a relative positive correction of 25\% with respect to the simulation 
is applied to the resolution.
A resolution of $4.28 \pm 0.19 \mevcc$ is obtained for the signal 
\sigmapmumu distribution and is used in the fit to define a Gaussian constraint
to the width of the signal distribution.
The combinatorial background is described as a modified ARGUS function with all 
parameters left free with the exception of the threshold, which is fixed to the 
kinematic limit. The shape of this background is also cross-checked with 
that of \pmumulfv candidates in data. 

\begin{figure}[tb]
\begin{center}
\includegraphics[width = \myfigurewidth]{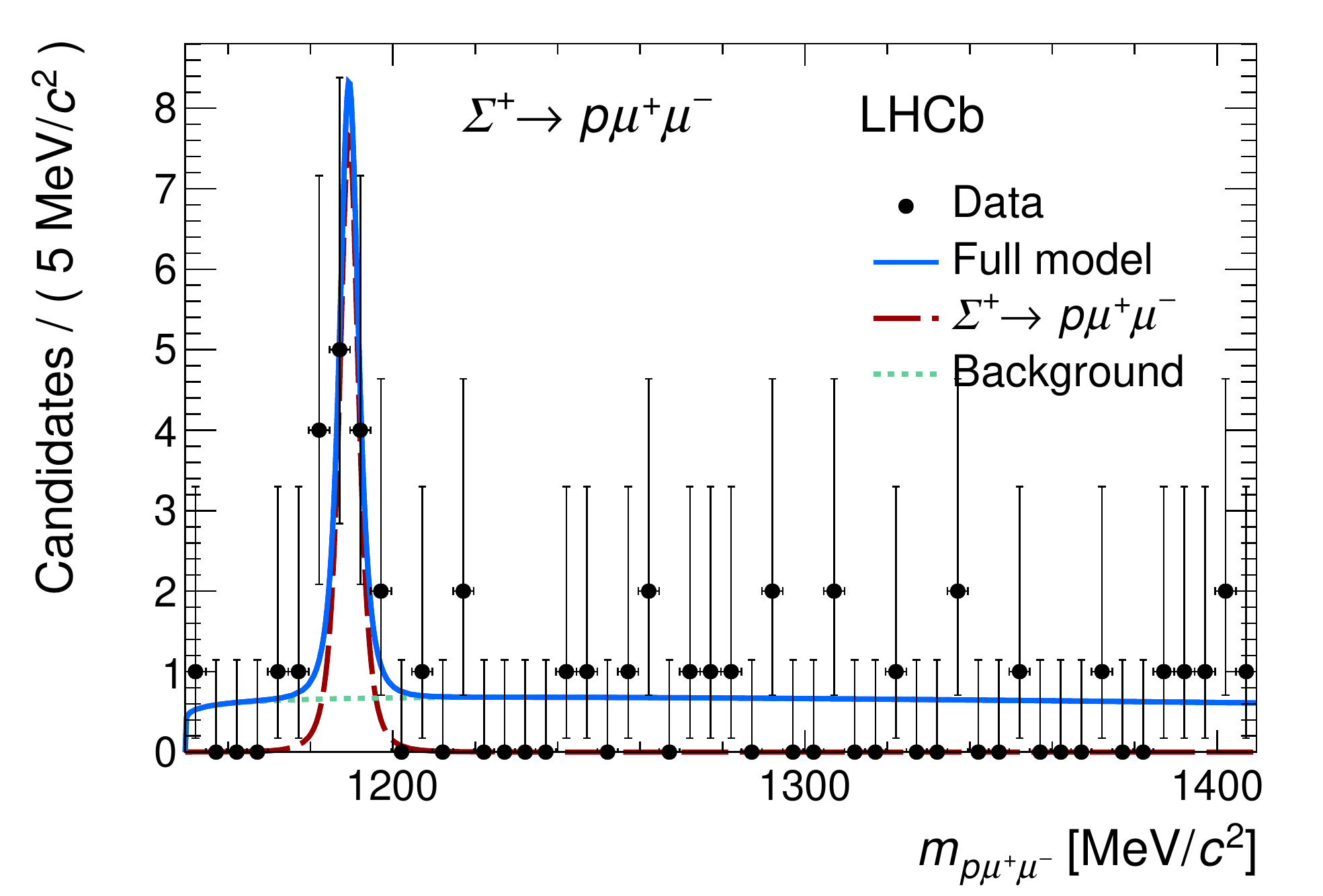}
\caption{Invariant mass distribution of \sigmapmumu candidates in data. }
\label{fig:invmass}
 \end{center}
\end{figure}

The invariant mass distribution of the \sigmapmumu candidates in data is shown
in Fig.~\ref{fig:invmass}. The significance of the signal is  
$\signdefault\,\sigma$, obtained from a comparison of the likelihood value of 
the nominal fit with that of a background-only fit~\cite{Wilks:1938dza}, 
and with the relevant systematic uncertainties included as Gaussian constraints to the
likelihood. A signal yield of \nsigmadefault is observed.
The corresponding branching fraction is ${\mathcal{B}(\sigmapmumu) = \brmeasuredsyst}$, 
where the first uncertainty is statistical and the second is systematic, 
consistent with the SM prediction. 
As a cross-check, the fit is repeated with tighter or looser requirements on the BDT 
or on the particle identification variables, and the signal yield is found to vary 
consistently with the signal efficiency. 
The fit is also repeated assuming a linear function for the background, in place of an ARGUS function, 
and the signal yield and significance are found to be stable.
Candidates in data are composed of about 48\% $\Sigmaplusbar$ anti-baryons in the final sample.

\begin{figure}[tb]
\begin{center}
  \includegraphics[width = \myfigurewidth]{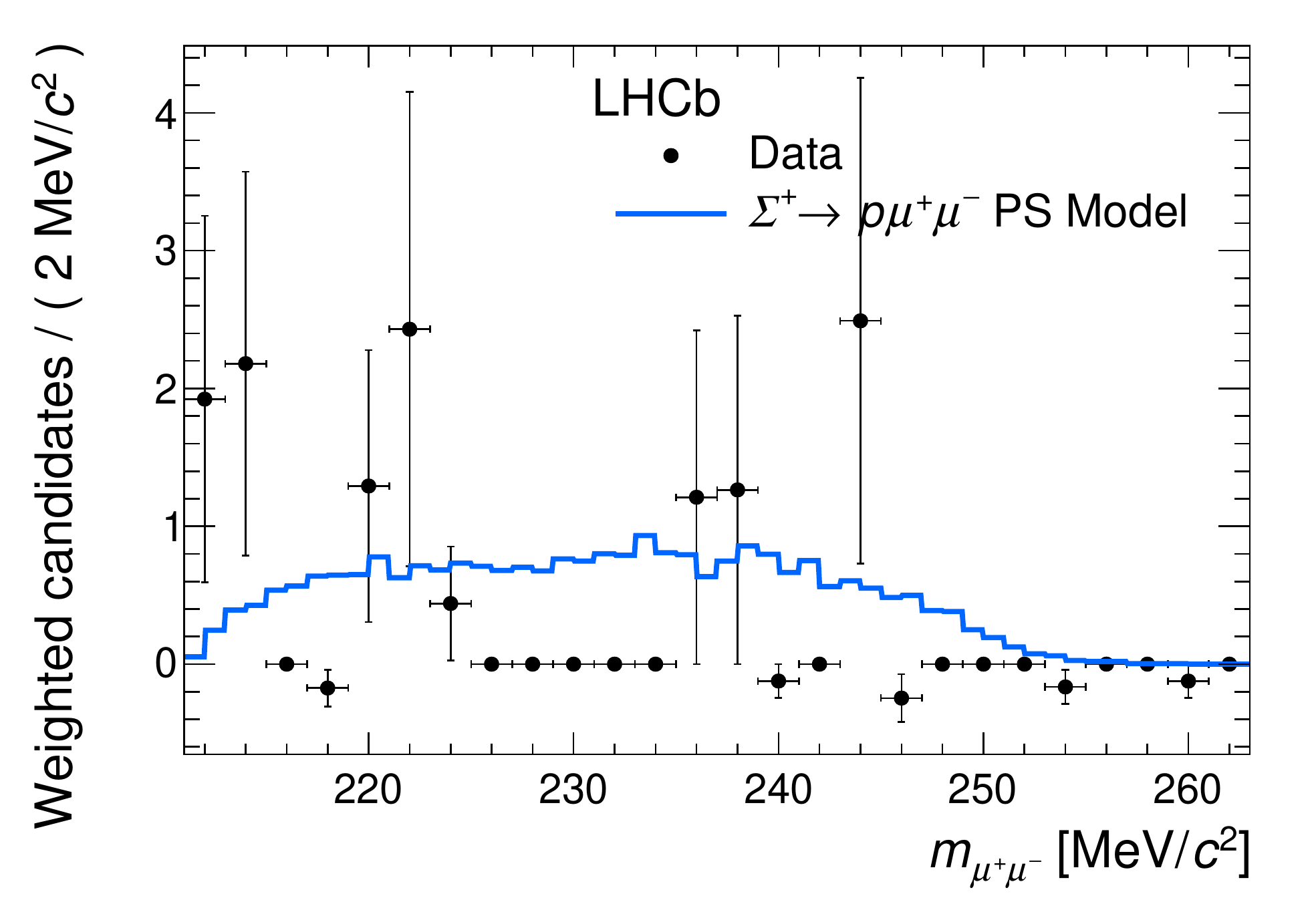}
\end{center}
  \caption{Background-subtracted distribution of the dimuon invariant mass for \sigmapmumu candidates, 
superimposed with the distribution from the simulated phase-space (PS) model. 
Uncertainties on data points are calculated as the square root of the sum of squared weights. }
  \label{fig:mumusplot}  
 \end{figure}

The distribution of the dimuon invariant mass
after background subtraction, performed with the \sPlot method~\cite{Pivk:2004ty}, is shown in Fig.~\ref{fig:mumusplot}.
A scan for a possible resonant structure in the
dimuon invariant mass is performed, considering a region within
two times the resolution in the \pmumu invariant mass around the known \Sigmaplus mass. 
The distribution of these candidates as a function of the dimuon invariant mass is shown in the supplemental material to this Letter~\cite{supplemental}.
Steps of half the resolution on the dimuon invariant mass, $\sigma(m_{\mu^+\mu^-})$, are considered in this scan, 
following the method outlined in Ref.~\cite{Williams:2015xfa}. 
The value of $\sigma(m_{\mu^+\mu^-})$ varies in the range [0.3, 2.3]\mevcc
depending on the dimuon invariant mass as shown in Ref.~\cite{supplemental}.
For each step the putative signal is estimated in a window of $\pm 1.5\times\sigma(m_{\mu^+\mu^-})$
around the considered particle mass,
 while the background is estimated from the lower and upper sidebands contained in the range $[1.5-4.0]\times\sigma(m_{\mu^+\mu^-})$
from the same mass. 
Only one of the two sidebands is considered when the second is outside the allowed kinematic range. 
The local p-value of the background-only hypothesis as a function of the dimuon mass is shown
 in Ref.~\cite{supplemental}, and no significant signal is found. 
The fit to the \pmumu invariant mass is then repeated restricting the sample to 
events within 1.5 times the resolution from the putative particle ($\mmumu \in [214.3\pm 0.75 ]\mevcc$).
No significant signal is found and a yield of \candidateshcp is measured corresponding to \fractionhcp of the \sigmapmumu yield. 
An upper limit on the branching fraction of the resonant channel is thus set with the $\rm{CL}_{\rm{S}}$ method~\cite{Read:2002hq} 
at $\mathcal{B}(\sigmapxmumu)<\upperlimithcpninety ~(\upperlimithcpninetyfive)$ at 90\% (95\%) confidence level.

In summary, a search for the \sigmapmumu rare decay is performed by the 
LHCb experiment using $pp$ collisions at centre-of-mass energies $\sqrt{s} = 7$ and $8 \tev$, 
corresponding to an integrated luminosity of $3 \invfb$. Evidence for the 
\sigmapmumu decay is found with a significance of $\signdefault$ standard deviations, 
including systematic uncertainties.
A branching fraction $\mathcal{B}(\sigmapmumu) = \brmeasured$ is measured, 
consistent with the SM prediction. 
No significant peak consistent with an intermediate particle is found 
in the dimuon invariant-mass distribution of the signal candidates.

\section*{Acknowledgements}
%
%
\noindent We express our gratitude to our colleagues in the CERN
accelerator departments for the excellent performance of the LHC. We
thank the technical and administrative staff at the LHCb
institutes. We acknowledge support from CERN and from the national
agencies: CAPES, CNPq, FAPERJ and FINEP (Brazil); MOST and NSFC
(China); CNRS/IN2P3 (France); BMBF, DFG and MPG (Germany); INFN
(Italy); NWO (The Netherlands); MNiSW and NCN (Poland); MEN/IFA
(Romania); MinES and FASO (Russia); MinECo (Spain); SNSF and SER
(Switzerland); NASU (Ukraine); STFC (United Kingdom); NSF (USA).  We
acknowledge the computing resources that are provided by CERN, IN2P3
(France), KIT and DESY (Germany), INFN (Italy), SURF (The
Netherlands), PIC (Spain), GridPP (United Kingdom), RRCKI and Yandex
LLC (Russia), CSCS (Switzerland), IFIN-HH (Romania), CBPF (Brazil),
PL-GRID (Poland) and OSC (USA). We are indebted to the communities
behind the multiple open-source software packages on which we depend.
Individual groups or members have received support from AvH Foundation
(Germany), EPLANET, Marie Sk\l{}odowska-Curie Actions and ERC
(European Union), ANR, Labex P2IO and OCEVU, and R\'{e}gion
Auvergne-Rh\^{o}ne-Alpes (France), RFBR, RSF and Yandex LLC (Russia),
GVA, XuntaGal and GENCAT (Spain), Herchel Smith Fund, the Royal
Society, the English-Speaking Union and the Leverhulme Trust (United
Kingdom).


\addcontentsline{toc}{section}{References}
\setboolean{inbibliography}{true}
\bibliographystyle{LHCb}
\bibliography{main,LHCb-PAPER,LHCb-CONF,LHCb-DP,LHCb-TDR}




 
\newpage
\centerline{\large\bf LHCb collaboration}
\begin{flushleft}
\small
R.~Aaij$^{40}$,
B.~Adeva$^{39}$,
M.~Adinolfi$^{48}$,
Z.~Ajaltouni$^{5}$,
S.~Akar$^{59}$,
J.~Albrecht$^{10}$,
F.~Alessio$^{40}$,
M.~Alexander$^{53}$,
A.~Alfonso~Albero$^{38}$,
S.~Ali$^{43}$
% ...

{\footnotesize \it
% ...
$ ^{5}$Clermont Universit{\'e}, Universit{\'e} Blaise Pascal, CNRS/IN2P3, LPC, Clermont-Ferrand, France\\
% ...
$ ^{10}$Fakult{\"a}t Physik, Technische Universit{\"a}t Dortmund, Dortmund, Germany\\
% ...
$ ^{38}$ICCUB, Universitat de Barcelona, Barcelona, Spain\\
$ ^{39}$Instituto Galego de F{\'\i}sica de Altas Enerx{\'\i}as (IGFAE), Universidade de Santiago de Compostela, Santiago de Compostela, Spain\\
$ ^{40}$European Organization for Nuclear Research (CERN), Geneva, Switzerland\\
% ...
$ ^{43}$Nikhef National Institute for Subatomic Physics, Amsterdam, The Netherlands\\
% ...
$ ^{48}$H.H. Wills Physics Laboratory, University of Bristol, Bristol, United Kingdom\\
% ...
$ ^{53}$School of Physics and Astronomy, University of Glasgow, Glasgow, United Kingdom\\
% ...
$ ^{59}$University of Cincinnati, Cincinnati, OH, United States\\
% ...
}
\end{flushleft}

\end{document}